\documentstyle[12pt,fleqn]{article}
\begin{document}
\vspace*{15mm}
\begin{center}
{\Large Evolution of Liouville density of a chaotic system} \\[2cm]
Asher Peres$^*$ and Daniel Terno\\[7mm]
{\sl Department of Physics, Technion---Israel Institute of
Technology, 32 000 Haifa, Israel}\\[1cm]
\end{center}\vfill

\noindent{\bf Abstract}\bigskip

An area-preserving map of the unit sphere, consisting of alternating
twists and turns, is mostly chaotic. A Liouville density on that sphere
is specified by means of its expansion into spherical harmonics. That
expansion initially necessitates only a finite number of basis functions.
As the dynamical mapping proceeds, it is found that the number of
non-negligible coefficients increases exponentially with the number of
steps. This is to be contrasted with the behavior of a Schr\"odinger
wave function which requires, for the analogous quantum system, a basis
of fixed size.\vfill

\vfill

\noindent $^*$\,Electronic address: peres@photon.technion.ac.il

\vfill\newpage

\begin{center}{\bf I. INTRODUCTION}\end{center}\bigskip

Chaos is commonly associated with nonlinear dynamics, and the
elusiveness of quantum chaos is sometimes attributed to the fact that
Schr\"odinger's equation is linear. However, the different behaviors of
classical and quantum systems cannot be explained so simply. Any
classical Hamiltonian motion can be described by means of a (possibly
singular) Liouville density, and the evolution of the latter obeys a
linear equation, which can be written as

\begin{equation} i\,\partial f/\partial t=Lf. \end{equation}
Here, $f$ is a function of  the $q^n$ and $p_n$, which are independent
variables parametrizing phase space, and $L$ is the Liouville operator,
or Liouvillian,

\begin{equation}
  L=\sum_n\,\biggl(\frac{\partial H}{\partial p_n}\biggr)\,\biggl(-i
   \,\frac{\partial}{\partial q^n}\biggr)-\biggl(\frac{\partial H}
  {\partial q^n}\biggr)\,\biggl(-i\,\frac{\partial}
  {\partial p_n}\biggr). \end{equation}
This operator is formally ``Hermitian'' (over a suitable domain of
Liouville functions $f$) so that the time evolution of $f$ is a {\em
unitary mapping\/} of phase space. Namely, if there is another Liouville
function, $g$, which also satisfies Eq.~(\theequation), the scalar
product \,\mbox{$\displaystyle\int$}$f^*g\,\prod dq^ndp_n$ \,is invariant in
time. This is Koopman's theorem [1].

The essential difference between the Liouville equation and the
Schr\"odinger equation is that, in the generic non-integrable case, the
spectrum of the Liouvillian is continuous~[2]. This gives rise to
fundamental differences between the evolution of Liouville densities and
that of quantum wave functions for bounded systems that have discrete
spectra in quantum theory. In particular, any quantum state can always be
represented, with arbitrary accuracy, by a {\em finite\/} number of
energy eigenstates. The time evolution of a bounded quantum system
therefore is multiply periodic, and will sooner or later have
recurrences [3]. On the other hand, the Liouville equation involves a
continuous spectrum, and the evolution of a Liouville density cannot be
represented by a finite number of terms.  Thanks to this property, a
Liouville density is able to become more and more convoluted with the
passage of time, and it may form intricate shapes with exceedingly thin
and long protuberances, something that a quantum wave function cannot do
[4].

The purpose of this work is to apply to the Liouville equation some of
the mathematical techniques that are standard in quantum mechanics. The
Liouville density $f$ is expanded into a complete set of orthonormal
functions,

\begin{equation} f(q,p)=\sum_n c_n\,u_n(q,p). \label{expf} \end{equation}
(Note that these $u_n$ are not the eigenfunctions of the
Liouvillian---the latter are not normalizable since its spectrum is
continuous.) The Liouville equation (1) then becomes

\begin{equation} i\,dc_m/dt=\sum_n L_{mn}\,c_n, \label{dcdt}\end{equation}
where $L_{mn}$ is a Hermitian matrix of infinite order. It follows that
$\sum|c_n|^2$ is constant in time. However, even if we start with a
small number of nonvanishing $c_n$, it turns out, as will be seen below,
that as time passes $f$ spreads over more and more $u_n$. A convenient
quantitative measure of this spread is the entropy-like expression

\begin{equation} S=-\sum_n |c_n|^2 \log |c_n|^2. \end{equation}
This $S$ is analogous to the ``entropy'' used in the study of quantum
chaos [5]. The intuitive meaning of $S$ is that $e^S$ is a rough
indication of the number of basis vectors that are appreciably involved
in the expansion of $f$ in Eq.~(\ref{expf}). An appropriate name for $S$
could be ``dimensional entropy'' (or $D$-entropy).

Since chaotic systems are characterized by the exponential divergence of
neighboring trajectories, we expect that the number of basis functions
needed for representing $f$ (with a given level of accuracy) should also
increase exponentially with time. In other words, we expect $S$ to
increase linearly with time for chaotic systems (in the asymptotic limit
of large $t$). On the other hand, for regular systems, whose
trajectories diverge linearly, we expect an asymptotically logarithmic
growth of $S$.

In the following sections, we shall investigate a simple dynamical model
that has a well documented chaotic behavior, and it will be seen that
these guesses are qualitatively correct. That model involves a discrete
time variable, rather than a continuous time, in order to make
calculations simpler. Instead of the continuous time evolution
(\ref{dcdt}), there is then a unitary transformation of the components
$c_n$ (explicitly given below).

\bigskip\begin{center}{\bf II. THE TWIST AND TURN MAP}\end{center}\bigskip

Consider a sequence of mappings of the unit sphere \,$x^2+y^2+z^2=1$, in
which each step consists of a {\it twist\/} by an angle $a$ around the
$z$-axis (namely, every $xy$ plane turns by an angle~$az$), followed by
a 90$^\circ$ rigid {\it rotation\/} around the $y$-axis. The result of
these consecutive twist and turn operations is

\begin{equation}\begin{array}{l}
 x'=z,\smallskip \\
 y'=x\,\sin(az)+y\,\cos(az),\smallskip \\
 z'=-x\,\cos(az)+y\,\sin(az).
\end{array} \label{TT} \end{equation}
This map is obviously area preserving. It was extensively investigated,
both classically and in quantum mechanics, by Haake, Ku\'s and Scharf
[6] who called it a ``kicked top.'' (It is not really like the motion of
a rigid top, because of the torsion.) For low values of~$a$, most
classical orbits are regular (that is, they are quasi-periodic). As $a$
increases, so does the fraction of chaotic orbits, until for $a=3$ most
of the sphere is visited by a single chaotic orbit, as may be seen in
Fig.~1. That figure also shows the presence of ``forbidden'' areas,
corresponding to regular regions located around fixed points of the map
[7]. All the following calculations refer to the case $a=3$, unless stated
otherwise.

The map (\ref{TT}) has interesting symmetries. Given any closed orbit,
another closed orbit can be obtained by changing the signs of both $x$
and $z$. This symmetry will be called $R_y$ (it is a rotation by
180$^\circ$ around the $y$-axis). Moreover, for any pair of distinct
orbits related by $R_y$, it is possible to obtain a third orbit (of
twice the length) by means of a 180$^\circ$ rotation of that pair of
orbits around the $x$-axis. These symmetries have important consequences
for the classification of Liouville densities, as will be seen in the
next Section.

A characteristic property of each orbit is its Lyapunov exponent. In
the present case, it may be defined as follows: consider an infinitesimal
circle drawn on the sphere around the initial point of a fiducial orbit.
As the map proceeds, this infinitesimal circle is deformed into an
infinitesimal ellipse, having the same area. The ellipse rotates and
stretches or contracts in an ``erratic'' way at each step. Let $a_n$ be
the length of its semi-major axis after $n$ steps. The Lyapunov exponent
(per step) is defined as

\begin{equation} \lambda=\lim_{n\to\infty} \frac{\log(a_n/a_0)}{n}\,.
 \label{Lyap}\end{equation}
Such a limit indeed exists for a generic chaotic orbit [8, 9]. In the
special case of a regular orbit, the pulsations of the ellipse are
bounded, and therefore $\lambda=0$.

We have computed in this way the Lyapunov exponents for $10^5$ orbits
with randomly chosen initial points. Each orbit was terminated after
$10^4$ steps (this usually happened when the orbit was regular), or, for
chaotic orbits, when the major axis of the ellipse exceeded $10^{16}$
(because of the inevitable loss of precision in any further
computation). The details of the calculations are given in Appendix A,
and the results are displayed in Fig.~2. About 14\% of the orbits are
regular. For the chaotic ones, we obtained:

\begin{equation} \lambda=
  0.346\pm 0.071\qquad\qquad{\rm(average \,\pm \,standard\ deviation)}.
\end{equation}
Ideally, $\lambda$ should have had a sharp value, the same for all
chaotic orbits. This dispersion free value could in principle have been
found by more sophisticated methods [10], but it was not needed for our
present purpose.\clearpage

\bigskip\begin{center}{\bf III. MAPPING OF LIOUVILLE DENSITIES}
\end{center}\bigskip

Instead of considering individual orbits, that is, mapping of points
into points, a more general approach, which gives new and instructive
insights, is the mapping of Liouville densities. Let us imagine that an
infinitesimal ``mass'' $\rho\,dA$ (which may be positive or negative) is
attached to each area $dA$. Let us further assume that this mass is
conserved during the twists and turns of the unit sphere, so that its
density $\rho$ obeys the linear law

\begin{equation} \rho'(x',y',z')=\rho(x,y,z), \end{equation}
by virtue of $dA'=dA$. Note that, as the mapping (\ref{TT}) is continuous,
nearby points are mapped into nearby points, so that a blob is always
mapped into a single blob (never into several disjoint blobs).

For the twist and turn map (\ref{TT}), these Liouville densities may
belong to three invariant symmetry classes, according to their behavior
under $R_y$ and $R_x$ (namely, 180$^\circ$ rotations around the $y$- and
$x$-axes, respectively). For example, if  $\rho= F(x^2,y^2,z^2,xyz)$ is
a single valued function of its four  arguments, this $\rho$ is even
under $R_y$ and $R_x$, and is mapped by Eq.~(\theequation) into another
function of the same type.  Likewise, any $\rho= yF(x^2,y^2,z^2,xyz)$ is
even under $R_y$ and odd under $R_x$, and is mapped by~(\theequation)
into a function of the same type. For instance, the function $\rho=y$
has this property, because

\begin{equation} y\to y'=x\,\sin(az)+y\,\cos(az)=y\left[\cos(az)+
 \frac{xyz}{y^2}\,\frac{\sin(az)}{z}\right]. \end{equation}

In general, any $\rho(x,y,z)$ can be written as the sum of three terms,
belonging to one the symmetry classes listed in Table~1.\clearpage

\vspace{2mm}\begin{center}
Table 1. \,Symmetry classes of Liouville functions.\\ \nopagebreak
\bigskip\begin{tabular}{|c|c|c|}\hline\hline
$R_y$ & $R_x$ & Functional form \rule{0pt}{2.6ex} \\ \hline
even & even & $\rho=F(x^2,y^2,z^2,xyz)$ \rule{0pt}{2.6ex}\\ \hline
even & odd & $\rho=yF(x^2,y^2,z^2,xyz)$ \rule{0pt}{2.6ex}\\ \hline
\multicolumn{1}{|c|}{odd} & \multicolumn{2}{c|}{$xF_1(x^2,y^2,z^2,xyz)
      +zF_2(x^2,y^2,z^2,xyz)$}  \rule{0pt}{2.6ex}\\ \hline\hline
\end{tabular}\end{center}\vspace{2mm}

A natural way of expanding Liouville functions on the unit sphere is the
use of spherical harmonics,

\begin{equation} \rho(\theta,\phi)=\sum_{lm} C_{lm}\,Y^m_l(\theta,\phi),
  \label{exprho} \end{equation}
where the angles $\theta$ and $\phi$ are related to the cartesian
coordinates in Eq.~(\ref{TT}) in the usual way:
$x=\sin\theta\,\cos\phi$, etc.

We shall use the common (but not universal) sign conventions [11, 12]

\begin{equation}\begin{array}{l}
  Y^m_l(\theta,\phi)=\biggl[\mbox{$\displaystyle\frac{2l+1}
  {4\pi}\,\frac{(l-m)!}{(l+m)!}$}
   \biggr]^{1/2}\,(-1)^m\,e^{im\phi}\,P^m_l(\cos\theta),\smallskip \\
  Y^{-m}_l(\theta,\phi)=(-1)^m\,Y_l^{m*}(\theta,\phi),\end{array}
\end{equation}
where $m\geq 0$ and $P_l^m(\cos\theta)$ are the associated Legendre
polynomials. The $Y_l^m(\theta,\phi)$ are othonormal, so that

\begin{equation}
  C_{lm}=\int Y_l^{m*}(\theta,\phi)\,\rho(\theta,\phi)\,d\Omega,
\end{equation}
where $d\Omega=\sin\theta\,d\theta\,d\phi$. In particular, the total
mass, namely $\int\rho d\Omega=\sqrt{4\pi}C_{00}$, is constant for our
area-preserving map. We shall henceforth ignore the trivial $C_{00}$
component and consider only the entropy of the other ones.

The nontrivial $C_{lm}$ transform as follows: during a twist, $\theta$
is constant, and

\begin{equation} \phi\to\phi'=\phi+a\,\cos\theta. \end{equation}
We thus have $\rho'(\theta,\phi+a\,\cos\theta)=\rho(\theta,\phi)$, or

\begin{equation} \rho'(\theta,\phi)=\rho(\theta,\phi-a\,\cos\theta),
\end{equation}
whence

\begin{equation} C'_{jm}=\int Y_j^{m*}(\theta,\phi)\,
   \rho(\theta,\phi-a\,\cos\theta)\,d\Omega, \end{equation}
because $d\Omega'=d\Omega$. Substitution of (\ref{exprho}) into
(\theequation) then gives

\begin{equation} C'_{jm}=\sum_l U^{(m)}_{jl}\,C_{lm}, \end{equation}
where the $U^{(m)}_{jl}$ are components of a unitary matrix:

\begin{equation}
 U^{(m)}_{jl}=\int Y_j^{m*}(\theta,\phi)\,Y_l^m(\theta,\phi)\,
     e^{-ima\cos\theta}\,d\Omega. \label{Ujl} \end{equation}
An accurate method for computing $U^{(m)}_{jl}$ for large $j$ and $l$ is
described in Appendix B.

We thus see that a twist leaves $m$ invariant but it introduces all the
$l\geq|m|$ (with exponentially small coefficients for large $l$), while
a rotation mixes different $m$ but leaves $l$ invariant. This emergence
of higher and higher $l$ occurs only in the {\it classical\/} twist and
turn map, which thus behaves in a quite different way from the
corresponding quantum map [6, 7]. This is because in quantum theory $l$
has the meaning of total angular momentum, and the latter is a constant
of the motion under a twist (which is generated by $J_z^{\,2}$), while
the classical $l$ has no such dynamical meaning and therefore need not
be conserved.

Note that a pure twist is a regular motion: all the orbits are closed
circles around the $z$-axis. We examined the growth of $S$, as a
function of a continuous twist angle $a$, for two maps starting with
$\rho$ proportional to $x$ (odd symmetry class) or to $y$ (even-odd
symmetry).  These are represented by initial states with

\begin{equation} C_{11}=\mp\,C_{1,-1}=1/\sqrt{2}, \end{equation}
respectively, and all other $C_{lm}=0$. Since different values of $m$ do
not mix during a twist, the $D$-entropy is the same in both cases.
Figure~3 shows the result: the growth of $e^S$ is roughly linear, as
expected.

We now turn our attention to the rotations of the unit sphere. With
spherical harmonics used as a basis, the matrix representation of a
90$^\circ$ rotation is well known [13, 14]: the index $l$ is not
affected, and for each $l$ the $(2l+1)$ components indexed by $m$
undergo a unitary transformation. Appendix B explains how to construct
accurately these unitary matrices (we proceeded up to $l=500$).

We checked the accuracy of our numerical calculations by verifying
that unitarity held at each step, with an error less than $10^{-8}$. To
achieve this result, we had to use a range of values of $l$ that
increased by more than a factor 3 at each step. We thus had, at the
fifth step, components $C_{lm}$ with $l$ up to 500. This implied that
the rotation matrices had all possible odd orders up to 1001, and twist
matrices had all orders up to 500. The next step would have exceeded the
capacity of our computer (or entailed a severe loss of accuracy).

Figure 4 shows how $S$ grows with the number of steps. In the case
$a=3$, we considered two different initial Liouville densities, given by
Eq.~(\theequation). These functions, which belong to different symmetry
classes, turn out to have roughly linear rates of growth of their
$D$-entropies, as expected, but, surprisingly, these rates are
manifestly different from each other. We must therefore conclude that
the growth of the $D$-entropy, contrary to the Kolmogorov-Sinai entropy
[8, 9], is not related in a simple way to the Lyapunov exponents of
individual orbits, because generic aperiodic orbits have no symmetry.

We also performed similar calculations for some lower values of $a$. For
$a=2$, the twist and turn map is mostly regular, but there still are
small chaotic regions [6]. These chaotic regions become almost invisible
for $a=1.5$ or less. We thus expected to find a growth of $S$ starting
about logarithmically, as for regular systems, and gradually becoming
linear, with a small slope, due to the presence of small chaotic
regions. However it turned out, as Fig.~4 shows, that for these low
values of $a$ the growth of $S$ is not uniform. Rather, $S$ oscillates
about a slowly increasing average. It is therefore hopeless to detect a
clear transition from the logarithmic regime to the linear one, and we
did not proceed further in these numerical simulations. (We also found
oscillations for $a=3$, when we started with an unsymmetrized $f$, for
example with $C_{11}=1$, and all other $C_{lm}=0$.)

\bigskip\begin{center}{\bf IV. CONCLUDING REMARKS}\end{center}\bigskip

We found some results that were not unexpected, but we also had several
surprises for which we can offer no explanation, and which may perhaps
be worth further investigation. There can be no doubt that, for a given
Liouvillian, the rate of growth of the $D$-entropy depends on the
symmetry class of the Liouville density. Therefore, contrary to the
Kolmogorov-Sinai entropy [8, 9], the $D$-entropy is not directly related
to the Lyapunov exponent of classical trajectories.

Even more surprising is a quantitative comparison of Figures 3 and 4.
The growth of $S$ for a pure twist (which is a regular mapping) is
{\it faster\/} than its growth for a weakly chaotic twist and turn map
(with low $a$), when we compare the total twist angle in the first case,
and the sum of discontinuous twists of the discrete map.

The initial motivation of this work was a search for the existence of
genuine quantum chaos (namely, phenomena that are not only a quantum
simulation of classically chaotic properties). We have found definite
clues that quantum chaos may indeed appear if, and only if, the quantum
system is unbounded and has a continuous spectrum.

\bigskip\begin{center}{\bf ACKNOWLEDGMENT}\end{center}\bigskip

This work was supported by the Gerard Swope Fund, and the Fund
for Encouragement of Research.\clearpage

\bigskip\begin{center}{\bf APPENDIX A. \,THE LYAPUNOV EXPONENT}
\end{center}\bigskip

Consider infinitesimal deviations $x\to x+\epsilon\xi$, \
$y\to y+\epsilon\eta$, and $z\to z+\epsilon\zeta$, from a fiducial orbit
of the map (\ref{TT}). These deviations satisfy

\begin{equation} x\xi+y\eta+z\zeta=0, \end{equation}
so as to lie on the surface of the sphere. We can also define the
tangential components of the vector $(\xi,\eta,\zeta)$, namely

\begin{equation}
  u=(-y\xi+x\eta)/\sqrt{1-z^2}\qquad\qquad{\rm and}\qquad\qquad
     v=\zeta/\sqrt{1-z^2}, \end{equation}
in the ``east'' and ``north'' directions, respectively.

Likewise, at the next step, let $x'\to x'+\epsilon\xi'$, etc. Neglecting
terms of order $\epsilon^2$ and higher, we obtain from Eq.~(\ref{TT}),

\begin{equation}\begin{array}{l}
 \xi'=\zeta,\smallskip \\
 \eta'=\xi\,\sin(az)+\eta\,\cos(az)-\zeta az',\smallskip \\
 \zeta'=-\xi\,\cos(az)+\eta\,\sin(az)+\zeta ay'.
\end{array} \label{tau} \end{equation}
This set of equations is the {\it linear\/} dynamical law for the
evolution of the vector $(\xi,\eta,\zeta)$. We now want to know how an
infinitesimal ``unit'' circle, namely a set of points with initial
components $u=\cos\alpha$, \,$v=\sin\alpha$ (where $\alpha$ runs from 0
to $2\pi$), transforms into an ellipse (with the same area). The
asymptotic growth of the major axis of this ellipse gives the Lyapunov
exponent, as defined by Eq.~(\ref{Lyap}).

To find how this infinitesimal circle transforms, we note that, by
virtue of the linearity of (\ref{tau}), if the initial components of a
tangential vector are ($\cos\alpha$) and ($\sin\alpha$), these
components become, after a number of steps,
$(T_{11}\cos\alpha+T_{12}\sin\alpha)$ and
$(T_{21}\cos\alpha+T_{22}\sin\alpha$), respectively, where the
coefficients $T_{pq}$ are independent of $\alpha$. It is therefore
enough to consider two infinitesimal tangent vectors having, initially,
$\alpha=0$ and $\pi/2$. Their evolution determines all the components of
$T_{pq}$, and the length of the semi-major axis is then obtained by
maximizing the expression

\begin{equation}  r^2(\alpha)=(T_{11}\cos\alpha+T_{12}\sin\alpha)^2+
 (T_{21}\cos\alpha+T_{22}\sin\alpha)^2, \end{equation}
as a function of $\alpha$. The result after $n$ steps is

\begin{equation} a_n=[\sigma+(\sigma^2-1)^{1/2}]^{1/2}, \end{equation}
where $\sigma=\sum_{pq}(T_{pq})^2$. Finally, the Lyapunov exponent is
given by Eq.~(\ref{Lyap}), in the limit $n\to\infty$.

\bigskip\begin{center}{\bf APPENDIX B. \,THE TWIST MATRIX}
\end{center}\bigskip

The evaluation of the matrix elements $U^{(m)}_{jl}$ in Eq.~(\ref{Ujl})
is the procedure which consumes most of the time in our calcualtions.
Each one of the indices, $j,\ l,\ m$, may run up to 500. There are
therefore many millions of different integrals for which no analytic
expression is known. These integrals are, for large values of their
indices, rapidly varying functions of $\cos\theta$, and a
straightforward numerical integration, with the level of accuracy that
we wanted, would be prohibitive.

We took advantage of the fact that a twist by a finite angle $a$ can be
generated by a sequence of infinitesimal twists by an angle $\epsilon$,
for which we can replace $e^{-im\epsilon z}$ by $(1-im\epsilon z)$. This
entails no loss of precision if $m\epsilon<10^{-16}$, when we compute
with 16 significant digits. The matrix elements $U^{(m)}_{jl}(\epsilon)$
can be evaluated explicitly (see below), and each matrix is then raised
to the appropriate power. For example, by taking $\epsilon=2^{-k}a$, we
merely have to compute $(\ldots((U^2)^2)^2\ldots)^2$, \,$k$ times.

We only explicitly need

\begin{equation}
 \int Y_j^{m*}(\theta,\phi)\,Y_l^m(\theta,\phi)\,\cos\theta\,d\Omega
 ={\rm const.}\times\int^1_{-1} P^m_j(z)\,P^m_l(z)\,z\,dz. \end{equation}
This expression is readily evaluated by using the identity

\begin{equation}
 z\,P^m_j(z)\equiv[(j-m+1)\,P^m_{j+1}(z)+(j+m)\,P^m_{j-1}] /(2j+1),
\end{equation}
and the orthogonality relations for associated Legendre polynomials [15].

\bigskip\begin{center}{\bf APPENDIX C. \,THE ROTATION MATRIX}
\end{center}\bigskip

There are explicit formulas for the unitary rotation matrices
$R^{(j)}_{mn}$ [13, 14] However, when $j$ is large, these formulas are
not convenient for numerical calculations, because each matrix element
is given as the tiny algebraic sum of a large number of huge terms with
opposite signs. Only a few elements with $m=\pm j$, or $m$ close to $\pm
j$, can easily be obtained from the general formula.

A much more efficient way of obtaining the $R$ matrix for a 90$^\circ$
turn around the $y$-axis is to use its very definition, namely
$R^\dagger J_x R= J_z$, or

\begin{equation} J_x\,R=R\,J_z. \end{equation}
With the standard representation, namely $(J_z)_{mn}=m\delta_{mn}$ and
$J_x$ real, this gives

\begin{equation} [(j+m)\,(j-m+1)]^{1/2}\,R^{(j)}_{m-1,n}+
 [(j+m+1)\,(j-m)]^{1/2}\,R^{(j)}_{m+1,n}=2n\,R^{(j)}_{mn}.
 \label{rec} \end{equation}

Let us therefore define, for each $j$ and $n$, a ``vector'' $V_m$ by

\begin{equation}
 R^{(j)}_{mn}=\frac{(-1)^{n-m}}{2^j}\,\biggl[\,\frac{(j+m)!\,
   (j-m)!}{(j+n)!\,(j-n)!}\,\biggl]^{1/2}\,V_m. \end{equation}
{}From the explicit formulas mentioned above [13, 14], it can be seen that
all the components of $V_m$ are integers, and in particular \,$V_{-j}=1$
\,and \,$V_{1-j}=2n$. The recursion relation (\ref{rec}) thus becomes

\begin{equation} (j-m+1)\,V_{m-1}-2n\,V_m+(j+m+1)\,V_{m+1}=0.
\end{equation}
(This equation also appeared as the last equation of ref.~[7],
unfortunately marred by a misprint. The calculations in ref.~[7] were
correct.)

Even with the simple recursion formula (\theequation), it is not trivial
to obtain all the $V_m$ for $j>50$ (when using double precision floating
point arithmetics), because the recursion is unstable and small
numerical errors grow exponentially. We therefore performed this
calculation {\it exactly\/}, using integer arithmetics; and we checked
that the resulting matrices were indeed unitary.

\newpage\frenchspacing
\begin{enumerate}
\item B. O. Koopman, Proc. Nat. Acad. Sci. {\bf 17}, 315 (1931).
\item B. O. Koopman and J. von Neumann, Proc. Nat. Acad. Sci. {\bf 18},
 255 (1932).
\item I. C. Percival, J. Math. Phys. {\bf 2}, 235 (1961).
\item H. J. Korsch and M. V. Berry, Physica D {\bf 3}, 627 (1981).
\item A. Peres, {\it Quantum Theory: Concepts and Methods\/} (Kluwer,
 Dordrecht, 1993) p.~360.
\item F. Haake, M. Ku\'s, and R. Scharf, Zeits. Phys. B {\bf 65}, 381
 (1987).
\item A. Peres, in {\it Quantum Chaos: Proceedings of the Adriatico
 Research Conference on
 Quantum Chaos\/}, ed. by H. A. Cerdeira, R. Ramaswamy, M. C. Gutzwiller,
 and G.~Casati (World Scientific, Singapore, 1991) pp.~73--102.
\item L. E. Reichl, {\it The Transition to Chaos\/} (Springer, New York,
1992) pp.~46--51.
\item M. Tabor, {\it Chaos and Integrability in Nonlinear Dynamics\/}
(Wiley, New York, 1989) pp.~148--151.
\item S. Habib and R. D. Ryne, Phys. Rev. Letters {\bf 74}, 70 (1995).
\item S. Gasiorowicz, {\it Quantum Physics\/} (Wiley, New York, 1974)
 p.~175.
\item E. Merzbacher, {\it Quantum Mechanics\/} (Wiley, New York, 1970)
 p.~185.
\item E. P. Wigner, {\it Group Theory and Its Application to the
 Quantum Mechanics of Atomic Spectra\/} (Academic Press, New York, 1959)
 p.~167.
\item M. Tinkham, {\it Group Theory and Quantum Mechanics\/}
 (McGraw-Hill, New York, 1964) p.~110.
\item M. Abramowicz and I. A. Stegun, {\it Handbook of Mathematical
 Functions\/} (Dover, New York, 1972) p.~334.
\end{enumerate}
\nonfrenchspacing\newpage

\parindent 0mm
{\bf Captions of figures}\bigskip

FIG. 1. \ Area-preserving projection of the hemisphere $y>0$. The figure
shows 20\,000 points belonging to a single chaotic orbit. The empty
regions are filled by regular orbits, not shown here.\bigskip

FIG. 2. \ Distribution of Lyapunov exponents for $10^5$ randomly chosen
orbits. Each bin of the histogram has width $10^{-2}$.\bigskip

FIG. 3. \ Growth of the $D$-entropy for a continuous twist, as a
function of the twist angle.\bigskip

FIG. 4. \ Growth of the $D$-entropy for the twist and turn map, for
various values of the twist parameter $a$.
\end{document}